\documentclass[11pt, superscriptaddress, nofootinbib,onecolumn]{revtex4-2}

\usepackage{tikz}

\usepackage{amssymb}
\usepackage{amsmath}
\usepackage{graphicx}
\usepackage{caption}
\usepackage{float}

\usepackage{natbib}
\begin{document}
	
	\newcommand{\dR}{\mathbb R}
	\newcommand{\dC}{\mathbb C}
	\newcommand{\dZ}{\mathbb Z}
	\newcommand{\id}{\mathbb I}
	\newtheorem{theorem}{Theorem}
	\newcommand{\ud}{\mathrm{d}}

	\title{Gauge-fixing and spacetime reconstruction in the Hamiltonian theory of cosmological perturbations}

	\author{Alice Boldrin}
	\email{Alice.Boldrin@ncbj.gov.pl}
	
	\author{Przemys\l aw Ma\l kiewicz}
	\email{Przemyslaw.Malkiewicz@ncbj.gov.pl}
	
	\affiliation{National Centre for Nuclear Research, Pasteura 7, 02-093 Warsaw, Poland} 
	
	\date{\today}
\begin{abstract}
We develop a complete Hamiltonian approach to the theory of perturbations around any spatially homogeneous spacetime. We employ the Dirac method for constrained systems which is well-suited to cosmological perturbations. We refine the method via the so-called Kucha\v r parametrization of the kinematical phase space. We separate the gauge-invariant dynamics of the three-surfaces  from the three-surface deformations induced by linear coordinate transformations. The canonical group of the three-surface deformations  and the complete space of gauge-fixing conditions are explicit in our approach. We introduce a frame in the space of gauge-fixing conditions and use it to considerably simplify the prescription for gauge-fixing, partial gauge-fixing and spacetime reconstruction. Finally, we illustrate our approach by considering the perturbed Kasner universe, for which we discuss two kinds of gauges that correspond respectively to the Coulomb-like and the Lorenz-like gauge in electrodynamics.
\end{abstract}


	\maketitle

\section{Introduction}
Cosmological perturbation theory (CPT) is widely assumed to be a working model of gravity at large cosmological scales. It is also used for constructing quantum theories of the origin of the primordial structure in the Universe. A notable example here is the inflationary model, but it is not the only one. The Hamiltonian formalism for cosmological perturbations, which is the subject of the present article, is a prerequisite for canonical quantizations of CPT. 

Obviously, the Hamiltonian formalism for cosmological perturbations has been derived before, at least in some simple background spacetime models such as the Friedmann universe \cite{Mukhanov:1990me,Langlois:1994ec, Dapor_2013, Malkiewicz_2019, Artigas_2022, Domenech:2017ems} or the Bianchi Type I model \cite{Uzan,Agullo2020,boldrin2021dirac}. Therefore, our goal in the present work is somewhat different. First of all, we keep our considerations general without specifying any particular background model until the final section where we study, as an example, the perturbed Kasner universe. Secondly, we study the complete Hamiltonian formalism that includes both a gauge-independent description of cosmological perturbations as well as the issues of  gauge-fixing, gauge transformations and spacetime reconstruction. To our best knowledge the latter elements of the Hamiltonian formalism have not been sufficiently studied in the literature. Thirdly, we achieve our goal by using and refining the Dirac method for constrained systems { \cite{Dirac:1964:LQM}} which, despite being well-suited to cosmological perturbations, has not been used for that purpose until very recently. { In reference \cite{Malkiewicz_2019} the Dirac method was applied to the perturbed flat Friedmann universe with a perfect fluid. The key steps on which this previous work was based are outlined in Sec. II of the present paper and supplemented with additional comments on the hypersurface deformations in the linearized theory. A significant improvement of the Dirac method is then obtained in Sec. III of the present paper with the use of a new parametrization of the ADM phase space. As we will show, the new variables make the basic elements of the Dirac method such as gauge-fixing or spacetime reconstruction much clearer, and also simplify the key formulas.}

The usual approach to CPT is to identify gauge-invariant variables in the configuration space and to study their and only their dynamics. The physical interpretation of the gauge-invariant variables is ambiguous and depends on the choice of the coordinate system for the spacetime. This choice is always made implicitly by assuming that some linear combinations of the components of the metric tensor vanish. The Friedmann-Robertson-Walker models have been studied extensively, and there are some well-known `canonical' gauge-fixing conditions that can be used for various purposes \cite{Stewart_1990,patrick-cpt,sasaki-cpt,Mukhanov:1990me,Malik:2008im}. Much less is known about valid gauges and the associated physical interpretations of gauge-invariant variables in anisotropic spacetimes. The Dirac method, as we show, provides a powerful tool for setting valid gauges and for reconstructing the spacetime metric from gauge-invariant variables in the Hamiltonian formalism. In this method, it is straightforward to show that the gauge transformations form an abelian group of translations { that act in the ADM phase space}, the space of all valid gauges becomes explicit, and the distinction between complete and partial gauge-fixing is very clear and analogous to electrodynamics. 

The physical motivation for the present work is two-fold. First, purely classical, to develop a tool for studying gauge-fixing conditions in homogeneous spacetimes both isotropic and anisotropic. It  is interesting to note that in anisotropic spacetimes there may be many representations of gravitational waves in contrast to isotropic universes where gravitational waves admit a unique representation in terms of transverse and traceless metric perturbations. In anisotropic spacetimes, as we show, a gravitational wave may be represented by a scalar metric perturbation. Second, to fully develop a relatively uncomplicated and thus practical laboratory for canonical quantizations of gravity. Such issues as the relation between the reduced and kinematical phase spaces and their respective quantizations, other quantization ambiguities, the quantum fate of diffeomorphism invariance and the time problem, semiclassical spacetime reconstruction and quantum-to-classical transition can be conveniently studied within our framework.

The outline of the article is as follows. In Section \ref{adaptDirac} we first discuss the general form of the Hamiltonian in cosmological perturbation theory, the algebra of constraints and their consistency with the dynamics. We apply the Dirac procedure to remove the constraints and obtain the reduced Hamiltonian dynamics for reduced variables. Next we rewrite the reduced Hamiltonian framework in terms of Dirac observables so that the issues of dynamics and physical spacetime reconstruction are to be solved separately at this point. In order to reconstruct the full spacetime metric, the stability of gauge-fixing conditions under the action of the full Hamiltonian is studied. In Section \ref{kuchar} we present an alternative phase space description analogous to the so-called Kucha\v r decomposition of the full geometrodynamics. In this new description the gauge transformations are given by a simple expression, which facilitates gauge-fixing and partial gauge-fixing, and simplifies the problem of spacetime reconstruction. In Sec. \ref{sectionkasner} we apply our formalism to the Kasner universe. We conclude in Sec. \ref{conclusion}. 
 
\section{Hamiltonian formalism and the gauge-fixing procedure}\label{adaptDirac}
In this section we briefly show how the Dirac procedure for constrained systems \cite{Dirac:1964:LQM} can be adapted to the Hamiltonian formulation of cosmological perturbation theory. We discuss the notions of Dirac observables, reduced variables, constrained and  physical dynamics, and the relations between them. We also discuss how the spacetime picture can be reconstructed from gauge-fixing conditions.

\subsection{Cosmological perturbations as a constrained system}

As cosmological background models we assume spatially homogeneous spacetimes with spatial coordinates fixed in such a way as to make the background shift vector components $N^i$ vanish. Also, the diffeomorphism constraints of the background model $\mathcal{H}^{(0)}_i=0$ must vanish trivially. Then the Arnowitt-Deser-Misner (ADM) Hamiltonian \cite{Arnowitt_2008} expanded to second order around such a homogenous spacetime model reads:	
	\begin{align}\label{Htot}
		\mathbb{H}=\int \left( N\mathcal{H}^{(0)}+N\mathcal{H}^{(2)}+\delta N^\mu\delta \mathcal{H}_\mu\right)\ud^3x,
	\end{align}
where $N$ is the zeroth order lapse function, $\delta N^\mu$ are the first order lapse and shift functions, $\mathcal{H}^{(0)}$ and $\mathcal{H}^{(2)}$ are respectively the zeroth-order constraint and the second order scalar Hamiltonian, and $\delta \mathcal{H}_\mu$ are linearized scalar and diffeomorphism constraints. This Hamiltonian is a function of the homogeneous three-metric $\bar{q}_{ij}$ and three-momentum  $\bar{\pi}^{ij}$, and the pure inhomogeneous perturbations of the three-metric  $\delta q_{ij}=q_{ij}-\bar{q}_{ij}$ and three-momentum  $\delta{\pi}^{ij}={\pi}^{ij}-\bar{\pi}^{ij}$. { The Hamiltonian system may include any number of matter fields but for clarity we make explicit use of the gravitational variables only. We emphasize, however, that all the following results apply equally well to more general set-ups. The total canonical structure can be then shown to be the sum of the homogeneous and inhomogeneous canonical structures (see, e.g. Eq. (18) of \cite{Langlois:1994ec}),
\begin{align*}
\{\bar{q}_{ij},\bar{\pi}^{kl}\}=\mathcal{V}_0^{-1}\delta_i^{(k}\delta_j^{l)},~~\{\delta q_{ij}(x),\delta{\pi}^{kl}(y)\}=\delta_i^{(k}\delta_j^{l)}\delta^3(x-y),
\end{align*}
with all the remaining Poisson brackets vanishing ($\mathcal{V}_0$ is the coordinate volume of the spatial section usually assumed to equal $1$ as in the example of Sec. \ref{sectionkasner}).} The interpretation of the terms in Eq. \eqref{Htot} is as follows. The zeroth-order constraint $\mathcal{H}^{(0)}$ generates time transformations in the homogeneous background spacetime while keeping the inhomogeneous fields fixed. The first-order constraints $\delta\mathcal{H}_\mu$ generate linearized transformations of the inhomogeneous spacetime while keeping the homogeneous background fixed. The second-order Hamiltonian $\mathcal{H}^{(2)}$ generates the dynamics of perturbations that must occur simultaneously with the dynamics of the homogeneous background generated by $\mathcal{H}^{(0)}$.

The Hamiltonian \eqref{Htot} defines a gauge system in the sense that the constraints $\delta \mathcal{H}_\mu$ are first-class up to first order. Specifically, at each spatial point, the algebra of the linearized constraints reads
\begin{align}\label{comms0}\begin{split}
\{\delta \mathcal{H}_i,\delta \mathcal{H}_j\}=0,&~~\{\delta \mathcal{H}_j,\delta \mathcal{H}_0\}=0.\end{split}
\end{align}
We note that the linearized constraints commute strongly at first order, { that is, when zeroth- and second-order outputs of the Poisson brackets are legitimately neglected}, and thus the group of gauge transformations that they generate for each spatial point must be abelian. This is true independently from any particular choice of background spacetime model\footnote{ Eq. \eqref{comms0} turns out to be completely general for perturbation theory around any homogenous background: since the homogeneous and inhomogeneous variables commute with each other, any nontrivial output of the Poisson bracket can be either zeroth-order (coming from the bracket of first-order variables) or second-order (coming from the bracket of zeroth-order variables).}. Furthermore, the constraints are dynamically stable on the constraint surface, namely,
\begin{align}\label{comms}\begin{split}
\{\int (\mathcal{H}^{(0)}+\mathcal{H}^{(2)}),\delta \mathcal{H}_0(x)\}=-\delta \mathcal{H}^i_{~,i}(x)\approx 0,~~\{\int (\mathcal{H}^{(0)}+\mathcal{H}^{(2)}),\delta \mathcal{H}_i(x)\}= 0\end{split}
\end{align}
where $\delta \mathcal{H}^i(x)=\bar{q}^{ij}\delta \mathcal{H}_j(x)$ { and the weak equality `$\approx$’ means `equal at the constraint surface’.} 

Eqs \eqref{comms0} and \eqref{comms} are a linearized version of the algebra of hypersurface deformations of canonical relativity \cite{10.2307/100497}. The full deformation algebra, and hence its linearization, are universal in the sense that they do not depend on any particular theory of gravity \cite{HOJMAN197688}.  We find it interesting to note that the abelianization of this algebra can also naturally occur for spherically symmetric hypersurface deformations \cite{PhysRevD.105.026017,universe8030184}.

\subsection{Gauge-fixing conditions}
From now on we shall focus on the unconstrained formulation of the dynamics of perturbations while omitting the separate problem of formulating the unconstrained dynamics of the background quantities. In order to remove the gauge freedom generated by the constraints $\delta\mathcal{H}_\mu$, we choose 4 gauge-fixing conditions denoted by $\delta c_{\mu}=0$, such that the commutation relations between the gauge-fixing functions and the linear constraints form an invertible matrix, that is,
\begin{align}\label{det}
\textrm{det}\{\delta c_{\nu},\delta\mathcal{H}_{\mu}\}\neq 0.
\end{align}
The gauge-fixing conditions $\delta c_{\mu}=0$ are solved by introducing a set of reduced canonical variables. Specifically, the reduction of the formalism is obtained by replacing the set of the 12 ADM perturbation variables $(\delta q_{ij},\delta \pi^{ij})$ with a reduced set of 4 physical variables denoted by $(\delta q^{phys}_{I},\delta \pi_{phys}^{I})$, where $I=1,2$ (see Sec. \ref{sfs} for an example of the reduction). The ADM variables are reduced to those 4 independent physical variables by virtue of 4 gauge-fixing conditions and 4 constraints. The physical variables are meant to form a canonical coordinate system on the submanifold in the kinematical phase space, on which the gauge-fixing functions and the constraints vanish. We call this submanifold the physical phase space. The canonical structure of the physical phase space is encoded in the Dirac bracket,
\begin{align}\label{DB}
\{\cdot,\cdot\}_D=\{\cdot,\cdot\} -\{\cdot,\delta \phi_{\mu}\}\{\delta \phi_{\mu},\delta \phi_{\nu}\}^{-1}\{\delta \phi_{\nu},\cdot\},
\end{align}
where $\delta\phi_{\mu}\in (\delta \mathcal{H}_1,\dots,\delta \mathcal{H}_4,\delta c_{1},\dots,\delta c_{4})$. Note that the Dirac bracket depends on the choice of the gauge-fixing conditions $(\delta c_{1},\dots,\delta c_{4})$. 

The second-order part of the Hamiltonian \eqref{Htot} is next expressed in terms of the physical variables yielding the reduced Hamiltonian:
\begin{align}\label{redH}
\left(N\mathcal{H}^{(2)}+\delta N^\mu\delta \mathcal{H}_\mu\right)\Big|_{\delta c_{\mu}=0=\delta\mathcal{H}_{\mu}}=N\mathcal{H}^{(2)}_{red}(\delta q^{phys}_I,\delta \pi_{phys}^I).
\end{align}

Given the reduced set of variables, the reduced Hamiltonian and the Dirac bracket, we find the reduced Hamilton equations to read (up to first order):
\begin{align}\label{redHE}\begin{split}
\frac{\ud}{\ud t}\delta {q}^{phys}_{I}=\{\delta q^{phys}_{I},\int \big( N\mathcal{H}^{(0)}+N\mathcal{H}^{(2)}_{red}\big)\ud^3x\}_D,\\
\frac{\ud}{\ud t}\delta {\pi}_{phys}^{I}=\{\delta \pi_{phys}^{I},\int \big( N\mathcal{H}^{(0)}+N\mathcal{H}^{(2)}_{red}\big)\ud^3x\}_D.\end{split}
\end{align}
Note that the term $\int N\mathcal{H}^{(0)}$ generates the dynamics of the background coefficients that in general are included in the definitions of $\delta q^{phys}_{I}$ and $\delta \pi_{phys}^{I}$.

{ Note that the four linearized constraints $\delta\mathcal{H}_\mu$ form a closed algebra that is trivial because the linearized constraints strongly commute. Hence, they must generate four translations in the ADM perturbation phase space. Since the perturbation variables are real variables with no restrictions on their ranges, the translations are unbounded in the phase space. In other words, the gauge orbits have the topology of $\mathbb{R}^4$ at each spatial point. To fix a point in $\mathbb{R}^4$ we need to impose the vanishing of four linear combinations of perturbation variables, $\delta c_{\mu}=0$ (they represent four 3-d planes that all cross each other at exactly one point in $\mathbb{R}^4$). Therefore, there are no obstructions to gauge-fixing: if the four conditions $\delta c_{\mu}=0$ fix a gauge locally, they must fix it globally.}

\subsection{Gauge-invariant description}\label{gid}

\begin{figure}[H]

	\begin{center}
	\includegraphics[width=0.8\textwidth]{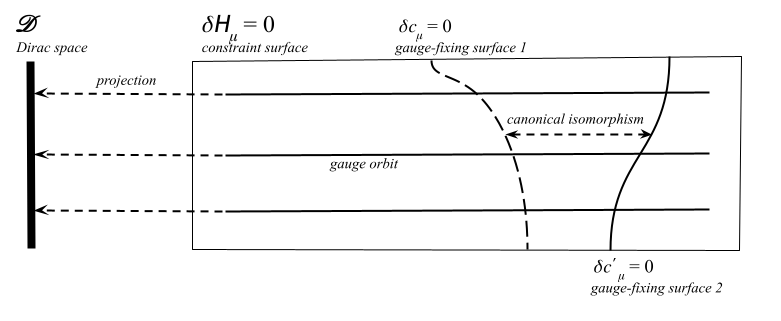}
	\end{center}
	\caption{Illustration of the key concepts involved in the Dirac procedure: the constraint surface, the gauge-fixing surface, the gauge orbit, the Dirac space and the canonical isomorphism between different gauge-fixing surfaces.}
	\label{fig:Dirac}
\end{figure}

Although obtained from a particular choice of gauge-fixing conditions, the reduced Hamiltonian and the physical variables in fact encode the gauge-independent dynamics of the model. This can be showed with the help of the Dirac observables, denoted by $\delta D_I$ and defined as follows:
\begin{align}\label{DDO}
\{\delta D_I,\delta \mathcal{H}_\mu\}\approx 0~~\textrm{for all}~\mu.
\end{align}
The Dirac observables commute with the 4 constraints $\delta\mathcal{H}_\mu$ and are understood as functions on the constraint surface. Hence, the number of independent Dirac observables must be equal to the number of the ADM perturbation variables minus 8 (4 constraints plus 4 gauge-fixing conditions), that is, equal to the number of the physical variables. The Dirac observables provide a parametrization of the space of the gauge orbits in the constraint surface whereas the physical variables provide a parameterization of a particular gauge-fixing surface that crosses each gauge orbit once and only once as depicted in Fig. \ref{fig:Dirac}. Therefore, there exists a one-to-one correspondence between the Dirac observables and the physical variables. Specifically, for any Dirac observable $\delta D_I$ there must exist a corresponding physical variable $\delta O^{phys}_I$ such that:
\begin{align}\label{diracbra}
\delta D_I+\xi_I^\mu\delta c_{\mu}+\zeta_I^\mu\delta\mathcal{H}_\mu=\delta O^{phys}_I(\delta q^{phys},\delta \pi_{phys}),
\end{align}
for some zeroth-order coefficients $\xi_I^\mu$ and $\zeta_I^\mu$. Moreover, this relation is a canonical isomorphism as (up to first order) 
\begin{align}\begin{split}
\{\delta D_I,\delta D_J\}=\{\delta D_I,\delta D_J\}_D&=\{\delta D_I+\xi_I^\mu\delta c_{\mu}+\zeta_I^\mu\delta\mathcal{H}_\mu,\delta D_J+\xi_J^\mu\delta c_{\mu}+\zeta_J^\mu\delta\mathcal{H}_\mu\}_D\\
&=\{\delta O^{phys}_I,\delta O^{phys}_J\}_D,
\end{split}\end{align}
{where the Dirac bracket (note the subscript `D') is defined in \eqref{DB}. A slightly more detailed discussion on this isomorphism can be found in \cite{Malkiewicz_2019}. Now,} the dynamics (\ref{redHE}) can be re-written in a gauge-independent manner as
\begin{align}\label{Dham}\begin{split}
\frac{\ud}{\ud t}\delta D_I=\{\delta D_I,\int \big( N\mathcal{H}^{(0)}+N\mathcal{H}^{(2)}_{red}\big)\ud^3x\}_D,\end{split}
\end{align}
where the reduced Hamiltonian is now understood as a function of the Dirac observables, $\mathcal{H}^{(2)}_{red}=\mathcal{H}^{(2)}_{red}(\delta D)$. The inclusion of the term $\{\delta D_I,\int N\mathcal{H}^{(0)}\ud^3x \}$ in Hamilton's equations \eqref{Dham} comes from the fact that the Dirac observables $\delta D_I$ are linear functions of the ADM perturbation variables $(\delta q_{ab},\delta \pi^{ab})$ with time-dependent background coefficients whose dynamics must also be taken into account. 

It is convenient to use the Dirac observables $\delta D_I$ (or, the physical variables $(\delta q^{phys}_{I},\delta \pi_{phys}^{I})$) as basic canonical variables. The respective canonical transformation is time-dependent at the level of perturbations and thus it generates an extra Hamiltonian density denoted by $\mathcal{H}^{(2)}_{ext}$. The new Hamiltonian density, called the physical Hamiltonian density, reads $\mathcal{H}^{(2)}_{phys}=\mathcal{H}^{(2)}_{red}+\mathcal{H}^{(2)}_{ext}$. With such a choice of basic variables the dynamics of the perturbations is now given purely by the second-order physical Hamiltonian:
\begin{align}\label{hphys}\begin{split}
\frac{\ud}{\ud t}\delta D_I=\{\delta D_I,\int N\mathcal{H}^{(2)}_{phys}\ud^3x\}_D,\end{split}
\end{align}
where the Dirac bracket  $\{A,B\}_D=\{A,\delta D_J\}\{\delta D_I,\delta D_J\}^{-1}\{\delta D_I,B\}$ can now be expressed in terms of  the Poisson bracket and the Dirac observables instead of the gauge-fixing functions. Note that this definition also depends on the gauge-fixing functions as the Dirac observables \eqref{DDO} are to some extent ambiguous and can be fixed by assuming that they commute with the gauge-fixing functions, i.e. $\{\delta D_I,\delta c_{\mu}\}_D=0$.

\subsection{Spacetime reconstruction}\label{ISR}

We have just seen that the problem of dynamics can be solved in a gauge-independent manner, that is, for all choices of gauge at once. Nevertheless, for the physical interpretation of the obtained dynamics the gauge-fixing conditions need to be specified. Thanks to the gauge-invariant formulation of the dynamics, the physical interpretation becomes a separate issue to be addressed independently of the dynamical equations. 

Thanks to the condition \eqref{det} there exists a one-to-one map between, on one hand, the values of the {gauge-fixing functions}, the {constraint functions} and the Dirac observables, and on the other hand, the values of the ADM perturbation variables, i.e.:
\begin{align}
(\delta\mathcal{H}_\mu,\delta c_{\mu},\delta D_I)\leftrightarrow (\delta q_{ab},\delta \pi^{ab}).
\end{align}
Thus, fixing $\delta\mathcal{H}_\mu=0$, $\delta c_{\mu}=0$ and assigning some numerical values to $\delta D_I$'s, unambiguously determines the geometry of the spatial leaf in terms of the ADM perturbation variables, in particular it determines the three-metric at any given time.

For the reconstruction of the full spacetime metric we still need to find the values of the first-order lapse and shift vector. We obtain them via the stability equation:
\begin{align}\label{coneq}
\{\delta c_{\nu},\mathbb{H}\}=0~~\Rightarrow~~\frac{\delta N^\mu}{N}=-\{\delta c_{\nu},\delta \mathcal{H}_\mu\}^{-1}\left(\{\delta c_{\nu},\mathcal{H}^{(0)}\}+\{\delta c_{\nu},\mathcal{H}^{(2)}\}\right).
\end{align}
The above equation is physically meaningful only in the constraint surface, that is, it holds weakly.

\section{Kucha\v r decomposition}\label{kuchar}
In the present section we revisit the procedure outlined above by means of the so-called Kucha\v r decomposition that is a special parametrization of the kinematical phase space with constraints encoded into canonical variables \cite{doi:10.1063/1.1666050, Hajicek:1999ht}. The existence of such a parametrization should become obvious as we proceed, nevertheless a general proof (i.e., valid beyond perturbation theory) can be found in \cite{Maskawa:1976hw}. The discussion of the choice of gauge-fixing conditions and the spacetime reconstruction turns out to be very transparent in this parametrization.

\subsection{Decomposition}
We start by introducing new canonical variables in the kinematical (ADM) phase space. Following Kucha\v r's decomposition we define two sets of canonical pairs. First we employ the constraints $\delta \mathcal{H}_\mu$ which, at first order, strongly commute between themselves (see Eq. \eqref{comms0}). Then we choose variables conjugate to the constraints, given by 4 gauge-fixing functions $\delta C^\mu$. These give us the first set of canonical pairs $(\delta \mathcal{H}_\mu,\delta C^\mu)$. Next, we define the so-called strong Dirac observables $\delta D_{I}$ that  are uniquely determined by the requirement that they strongly commute with the constraint functions and the gauge-fixing functions. Their Poisson algebra is closed and they form canonical pairs which we shall denote by $(\delta Q_I,\delta P^I)$. Alternatively, one might first choose the strong Dirac observables $(\delta Q_I,\delta P^I)$, which in turn would determine the respective gauge-fixing functions. 

Finally, the new set of canonical variables in the kinematical phase space reads:
\begin{equation*}(\delta \mathcal{H}_\mu,\delta C^{\mu},\delta Q_I,\delta P^I).\end{equation*}
It follows that, up to first order,
\begin{align}\label{PbD}\{\delta\mathcal{H}_\mu(x),\delta C^{\nu}(y)\}=\delta_{\mu}^{~\nu}\delta^3(x-y),~\{\delta Q_I(x),\delta P^J(y)\}=\delta_{I}^{~J}\delta^3(x-y),\end{align}
with all the remaining basic commutation relations vanishing. The canonical transformation
\begin{align}\label{Kmap}
\mathbb{R}^{12}\ni (\delta q_{ab},\delta \pi^{ab})\mapsto (\delta \mathcal{H}_\mu,\delta C^{\mu},\delta Q_I,\delta P^I)\in\mathbb{R}^{12}\end{align}
is time-dependent as the Kucha\v r variables are linear combinations of the ADM perturbation variables with time-dependent zeroth-order coefficients. Hence we need to determine the associated extra Hamiltonian $\mathbb{K}$ that must be included in the new Hamiltonian $\mathbb{H}_K$:
\[\mathbb{H}\rightarrow\mathbb{H}_K=\mathbb{H}+\mathbb{K}.\]
The extra Hamiltonian $\mathbb{K}$ is needed to compensate for the dynamics of the zeroth-order coefficients present in the definition of the new canonical variables. Notice that the Poisson brackets expressed in terms of the ADM variables and the Kucha\v r variables are equivalent up to first order only, and the commutation relations \eqref{PbD} are now assumed to be exact. Thus the model, when expressed in terms of the Kucha\v r variables, becomes an exact gauge system.

In the Kucha\v r parametrization the total Hamiltonian is given by 
	\begin{align}\label{HK}
		\mathbb{H}_K=\int\left(N\mathcal{H}^{(0)}+N(\mathcal{H}^{(2)}+\mathcal{K}^{(2)})+\delta N^\mu\delta \mathcal{H}_\mu\right)\ud^3x,
	\end{align}
where $\int N\mathcal{K}^{(2)}\ud^3x=\mathbb{K}$. It generates the following Hamilton equations:
	\begin{align}\label{eom}\begin{split}
	\delta \dot{Q}_I&=N\frac{\partial(\mathcal{H}^{(2)}+\mathcal{K}^{(2)})}{\partial \delta{P}^I},~~\delta \dot{P}^I=-N\frac{\partial(\mathcal{H}^{(2)}+\mathcal{K}^{(2)})}{\partial \delta{Q}_I},\\
	\delta \dot{\mathcal{H}}_\mu&=N\frac{\partial(\mathcal{H}^{(2)}+\mathcal{K}^{(2)})}{\partial \delta{C}^\mu},~~\delta \dot{C}^\mu=-N\frac{\partial(\mathcal{H}^{(2)}+\mathcal{K}^{(2)})}{\partial {\delta\mathcal{H}}_\mu}
	-\delta N^\mu.
	\end{split}
	\end{align}
The dynamical equations for the new canonical variables allow us to restrict the form of the Hamiltonian $\mathbb{H}_K$. Because the dynamics of the constraints is conserved in the constraint surface, the terms $\propto \delta{C}^\nu\delta{C}^\mu$, $\propto \delta Q_I\delta{C}^\mu$ and $\propto \delta P^I\delta{C}^\mu$ must be absent in $\mathbb{H}_K$. The last two terms must also vanish  by virtue of  the fact that the dynamics of the Dirac observables in the constraint surface must be independent of the choice of gauge. Moreover, it is clear that the Hamiltonian density $\mathcal{H}^{(2)}+\mathcal{K}^{(2)}$ must be weakly equal to the Hamiltonian density $\mathcal{H}^{(2)}_{phys}$ of Eq. \eqref{hphys}, more specifically, $\mathcal{H}^{(2)}\approx\mathcal{H}^{(2)}_{red}$ and $\mathcal{K}^{(2)}\approx\mathcal{H}^{(2)}_{ext}$. The latter term compensates for the dynamics of the background coefficients in the Dirac observables expressed in terms of the ADM perturbation variables.

Thus, the total Hamiltonian  $\mathbb{H}_K$ is made of the physical part  \eqref{hphys} and a weakly vanishing part. In the Kucha\v r parametrization the total Hamiltonian reads:
\begin{align}\label{KHw}
\mathbb{H}_K=N\int\left(\underbrace{\mathcal{H}^{(2)}_{phys}(\delta Q_I,\delta P^I)}_\text{physical part}+\underbrace{(\lambda_{1}^{\mu I}\delta{Q}_I+\lambda_{2I}^{\mu}\delta{P}^I+\lambda_{3}^{\mu\nu}\delta\mathcal{H}_\nu+\lambda_{4\nu}^\mu\delta C^\nu+\frac{\delta N^\mu}{N})\delta \mathcal{H}_\mu}_\text{weakly vanishing part}\right)\ud^3x,
\end{align}
where, in general, the zeroth-order coefficients $\lambda_1$, $\lambda_2$ and $\lambda_3$ depend on the particular choice of gauge-fixing functions $\delta C^{\mu}$. The gauge-dependence of  $\lambda_1$ and $\lambda_2$ becomes evident from \eqref{eom} after writing down the formula for the lapse function and shift vector which themselves are gauge-dependent:
\begin{align}\label{lapseshiftfull}
\frac{\delta N^\mu}{N}\approx -\lambda_{1}^{\mu I}\delta{Q}_I-\lambda_{2I}^{\mu}\delta{P}^I.
\end{align}
The value of $\lambda_3$ is irrelevant for the physical content of the theory. The value of $\lambda_4$ is gauge-invariant, that is, it does not depend on the particular gauge-fixing functions that are used. Indeed, we note from \eqref{KHw} that
$$\lambda_{4\nu}^\mu\delta \mathcal{H}_\mu=\{\delta\mathcal{H}_\nu,\mathbb{H}_K\}=\{\delta\mathcal{H}_\nu,\int (\mathcal{H}^{(0)}+\mathcal{H}^{(2)})\},$$
that is, the matrix $\lambda_4$ is fixed unambiguously by the algebra of hypersurface deformations \eqref{comms}.


\subsection{Gauge transformations}	
	The Kucha\v r decomposition provides a class of parametrizations of the kinematical phase space rather than a single, fixed parametrization. The relevant freedom in defining Kucha\v r's variables comes from the free choice of gauge-fixing functions. Let us use the twiddle mark over the new canonical quantities and, in particular, let {$\delta \tilde{C}^{\mu}$} denote a new set of gauge-fixing functions. The full gauge transformation is given by the canonical map $\mathbf{G}$:
\begin{align}\label{Kmap}
\mathbf{G}:~(\delta \mathcal{H}_\mu,\delta C^{\mu},\delta Q_I,\delta P^I)\mapsto (\delta \tilde{\mathcal{H}}_\mu,\delta \tilde{C}^{\mu},\delta \tilde{Q}_I,\delta \tilde{P}^I),\end{align} 
where $\delta \tilde{\mathcal{H}}_\mu=\delta \mathcal{H}_\mu$, i.e., the constraint functions are preserved by the map.
	
	We assume the new gauge-fixing functions $\delta \tilde{C}^{\mu}$ to be canonically conjugate to the constraints $\delta{\mathcal{H}}_\mu$, that is, $\{\delta\mathcal{H}_\mu(x),\delta \tilde{C}^{\nu}(y)\}=\delta_{\mu}^{~\nu}\delta^3(x-y)$. If this is not the case, there is a simple way to bring any gauge-fixing functions, say $\delta {C}^{\mu}_{ini}$, to the canonical form, say $\delta {C}_{can}$. Namely,
\begin{align}\label{can}
	\delta {C}_{can}^{\mu}(x)=\int M^\mu_{~\nu}(x,y)\delta {C}^{\nu}_{ini}(y)\ud^3y,~~\textrm{where}~~ M^{\mu}_{~\nu}(x,y)=\{\delta\mathcal{H}_\mu(x),\delta {C}^{\nu}_{ini}(y)\}^{-1}.
	\end{align}
Now, it is clear that the difference between the gauge-fixing functions should satisfy
\[\{\delta\mathcal{H}_\nu,\delta \tilde{C}^{\mu}-\delta {C}^{\mu}\}=0,\]
which has the solution
\begin{align}\label{newgauge}
\delta \tilde{C}^{\mu}=\delta C^\mu +\alpha^\mu_{~I}\delta P^I+\beta^{\mu I} \delta Q_{I}+\gamma^{\mu\nu} \delta\mathcal{H}_\nu,
\end{align}
where $\alpha^\mu_{~I}$, $\beta^{\mu I}$ and $\gamma^{\mu\nu}$ are background-dependent parameters. The first two parameters are\footnote{The index $I$ runs from $1$ to half of the number of basic Dirac observables for a given system. Thus, in the vacuum case $I\in\{1,2\}$ labels two polarization modes of the gravitational wave.} $4\times 2$ matrices, whereas $\gamma^{\mu\nu}$ is a $4\times 4$ matrix. Since gauge-fixing conditions are physically relevant only in the constraint surface it follows that the only independent parameters involved in the gauge transformation \eqref{newgauge} must be $\alpha^\mu_{~I}$ and $\beta^{\mu I}$. In other words, the space of gauge-fixing conditions for any fixed label $\mu$ is the affine space of dimension equal to the number of Dirac observables in the system. Note that the choice of gauge-fixing conditions can in principle depend on time through the time-dependent background quantities $\alpha^\mu_{~I}$ and $\beta^{\mu I}$. These parameters can be obtained with the formulas
\[
	\alpha^\mu_{~I}
	=
	\{\delta Q_I,\int(\delta \tilde{C}^\mu
	-
	\delta C^\mu)
	\},~~	
	\beta^{\mu I} 
	=
	\{\int(\delta \tilde{C}^\mu
	-
	\delta C^\mu)
	,\delta P^I
	\},
\]
where the Poisson bracket is given in any (i.e., the ADM or the Kucha\v r) parametrization.

The symplectic form,
\begin{align}
\Omega=d\delta Q_I\wedge d\delta P^I+ d \delta \mathcal{H}_\mu \wedge  d\delta C^\mu,
\end{align}
may be re-expressed as
\begin{align}
	\Omega
	&	
	=	
	d\left(\delta Q_I{-}{\alpha^\mu_{~I}\delta\mathcal{H}_\mu}\right)
	\wedge 
	d\left(\delta P^I{+}{\beta^{\nu I}\delta\mathcal{H}_\nu}\right)
	+
	d \delta\mathcal{H}_\mu
	\wedge
	d
	\left[
	\delta {C}^\mu+\alpha^\mu_{~I}\delta P^I
	 +\beta^{\mu I} \delta Q_{I}
	+
		\gamma^{\mu\nu}
	\delta\mathcal{H}_\nu
	\right] 
			\nonumber\\&
	+
	dt
	\wedge 
	d\left[
	\dot{\alpha}^\mu_{~I} \delta\mathcal{H}_\mu\delta P^I
	+\dot{\beta}^{\mu I}\delta\mathcal{H}_\mu \delta Q_{I}+\frac{1}{2}
	\left(
	\dot{\alpha}^\mu_{~I}
	\beta^{\nu I}
	-
	\alpha^\mu_{~I}
	\dot{\beta}^{\nu I}
	\right)
	\delta\mathcal{H}_\mu
	\delta\mathcal{H}_\nu
	\right],
\end{align}
with $\gamma^{\mu\nu}=\frac{1}{2}\left(\alpha^\mu_{~I}\beta^{\nu I}-\alpha^\nu_{~I}\beta^{\mu I}\right)$, leading to the new Kucha\v r variables,
\begin{align}\label{gtr}\begin{split}
	\delta\tilde{Q}_I&= \delta Q_I-\alpha^\mu_{~I}\delta\mathcal{H}_\mu,\\
	\delta\tilde{P}^I&=
	 \delta P^I+\beta^{\mu I}\delta\mathcal{H}_\mu,\\
	 \delta\tilde{\mathcal{H}}_\mu&=\delta\mathcal{H}_\mu,\\
	 \delta \tilde{C}^\mu
	&=
	\delta C^\mu
	+\alpha^\mu_{~I}\delta P^I
	 +\beta^{\mu I} \delta Q_{I}
	+
		{\gamma^{\mu\nu}}
\delta\mathcal{H}_\nu,
	 \end{split}
\end{align}
and the extra Hamiltonian density,
\begin{align}\label{Kden}
\Delta\mathcal{K}^{(2)}=-	\left[\dot{\alpha}^\mu_{~I} \delta\mathcal{H}_\mu\delta P^I
	+\dot{\beta}^{\mu I}\delta\mathcal{H}_\mu \delta Q_{I}+\frac{1}{2}
	\left(
	\dot{\alpha}^\mu_{~I}
	\beta^{\nu I}
	-
	\alpha^\mu_{~I}
	\dot{\beta}^{\nu I}
	\right)
	\delta\mathcal{H}_\mu
	\delta\mathcal{H}_\nu\right]
,
\end{align}
which, when added to \eqref{HK}, yields a new Hamiltonian that we shall denote by $\mathbb{H}_{\tilde{K}}$. The matrix $\gamma^{\mu\nu}$ is determined by $\alpha^\mu_{~I}$ and $\beta^{\mu I}$, as expected. Therefore, the parameters $\alpha^\mu_{~I}$ and $\beta^{\mu I}$ determine a complete gauge transformation. We note that the extra Hamiltonian is weakly zero as it must be in order for the dynamical equations \eqref{eom} for the Dirac observables to be preserved in the constraint surface. The gauge transformation does not affect the definition of the Dirac observables in the constraint surface, nevertheless it does modify their (physically irrelevant) extension beyond the constraint surface. 

From \eqref{gtr} we conclude that the local space of gauge-fixing conditions is  an affine space of dimension $n$ and the local gauge group is the space of displacement vectors in this affine space, $\mathbf{G}=\mathbb{R}^n$, where $n$ is the number of $\alpha^\mu_{~I}$'s and $\beta^{\mu I}$'s, that is, the number of Dirac observables times the number of gauge-fixing conditions. Hence, the group of canonical transformations \eqref{Kmap} is abelian,
\begin{align}
\mathbf{G}_{\alpha,\beta}\circ\mathbf{G}_{\alpha',\beta'}=\mathbf{G}_{\alpha+\alpha',\beta+\beta'}.
\end{align}
The physical part of the Hamiltonian \eqref{KHw} is transformed according to the following replacement: 
\[\mathcal{H}^{(2)}_{phys}(\delta Q_I,\delta P^I)\rightarrow \mathcal{H}^{(2)}_{phys}(\delta \tilde{Q}_I,\delta \tilde{P}^I),\] 
i.e., it is gauge-invariant. The weakly vanishing part of the Hamiltonian \eqref{KHw} is transformed as follows:
\begin{align}\label{ltr}\begin{split}
\lambda_{1}^{\mu I}\rightarrow \tilde{\lambda}_{1}^{\mu I}&=\lambda_{1}^{\mu I}-\dot{\beta}^{\mu I}-\lambda_{4\nu}^\mu\beta^{\nu I}-\frac{\partial^2\mathcal{H}^{(2)}_{phys} }{\partial \delta Q_I\partial \delta P^J}\beta^{\mu J}+\frac{\partial^2\mathcal{H}^{(2)}_{phys} }{\partial \delta Q_I\partial \delta Q_J}\alpha^\mu_{~J},\\
\lambda_{2I}^{\mu}\rightarrow \tilde{\lambda}_{2I}^{\mu}&=\lambda_{2I}^{\mu}-\dot{\alpha}^\mu_{~I}-\lambda_{4\nu}^\mu\alpha^\nu_{~I}+\frac{\partial^2\mathcal{H}^{(2)}_{phys} }{\partial \delta P^I\partial \delta Q_J}\alpha^\mu_{~J}-\frac{\partial^2\mathcal{H}^{(2)}_{phys} }{\partial \delta P^I\partial \delta P^J}\beta^{J\mu},\\
\lambda_{3}^{\mu\nu}\rightarrow \tilde{\lambda}_{3}^{\mu\nu}&=\lambda_{3}^{\mu\nu}+\frac{1}{2}\left(\dot{\alpha}^\mu_{~I}\beta^{\nu I}-\alpha^\mu_{~I}\dot{\beta}^{\nu I}\right)+\frac{1}{2}\lambda_{4\kappa}^\mu\left(\alpha^\kappa_{~I}\beta^{\nu I}-\alpha^\nu_{~I}\beta^{\kappa I}	\right)
+\lambda_{1}^{\mu I}\alpha^\nu_{~I}-\lambda_{2I}^{\mu}\beta^{\nu I}\\
&-\frac{\partial^2\mathcal{H}^{(2)}_{phys} }{\partial \delta Q_I\partial \delta P^J}\alpha^\mu_{~I}\beta^{\nu J}
+\frac{1}{2}\frac{\partial^2\mathcal{H}^{(2)}_{phys} }{\partial \delta Q_I\partial \delta Q_J}\alpha^\mu_{~I}\alpha^\nu_{~J}
+\frac{1}{2}\frac{\partial^2\mathcal{H}^{(2)}_{phys} }{\partial \delta P^I\partial \delta P^J}\beta^{\mu I}\beta^{\nu J},\\
\lambda_{4\nu}^\mu\rightarrow \tilde{\lambda}_{4\nu}^\mu&=\lambda_{4\nu}^\mu,
\end{split}\end{align}
i.e., $\lambda_{1}$, $\lambda_{2}$ and $\lambda_{3}$ are gauge-dependent, whereas $\lambda_{4}$ is gauge-invariant, as previously mentioned.

\subsection{Spacetime reconstruction}

The gauge stability condition $\delta \dot{C}^\nu= 0$ is a dynamical equation. Therefore we need to be clear on which particular parametrization we use in the definition of the Poisson bracket so that the correct Hamiltonian is used in the dynamical equation. We shall denote the Poisson bracket in the Kucha\v r parametrization by $\{\cdot,\cdot\}_K$. Thus, the gauge stability condition reads:
\begin{align}
\{\delta C^\nu, \mathbb{H}_{K}\}_K=0,
\end{align}
or, making use of Eq. \eqref{HK},
\begin{align}\label{deltaN}
	\frac{\delta N^\mu}{N}
	=
	-\frac{\partial(\mathcal{H}^{(2)}+\mathcal{K}^{(2)})}{\partial {\delta\mathcal{H}}_\mu}.
\end{align}
The above formula involves only the weakly vanishing part of the Hamiltonian \eqref{KHw} as the lapse and shift are pure gauge-dependent quantities. However, the difference between those quantities for two different gauges depends only on the gauge-independent part of the Hamiltonian  \eqref{KHw}, which simplifies the task of spacetime reconstruction. Indeed, after substituting  Eq. \eqref{ltr} into Eq. \eqref{lapseshiftfull} we find that in the constraint surface
\begin{align}\label{lapseshift}\begin{split}
\frac{\delta\tilde{N}^\mu}{N}\bigg|_{\delta\tilde{C}^{\mu}=0}- \frac{\delta{N}^\mu}{N}\bigg|_{\delta{C}^{\mu}=0}&\approx \left(\lambda_{4\nu}^\mu\beta^{\nu I}+\dot{\beta}^{\mu I}+\frac{\partial^2\mathcal{H}^{(2)}_{phys} }{\partial \delta Q_I\partial \delta P^J}\beta^{\mu J}-\frac{\partial^2\mathcal{H}^{(2)}_{phys} }{\partial \delta Q_I\partial \delta Q_J}\alpha^\mu_{~J}\right)\delta{Q}_I\\
&
+\left(\lambda_{4\nu}^\mu\alpha^\nu_{~I}+\dot{\alpha}^\mu_{~I}-\frac{\partial^2\mathcal{H}^{(2)}_{phys} }{\partial \delta P^I\partial \delta Q_J}\alpha^\mu_{~J}+\frac{\partial^2\mathcal{H}^{(2)}_{phys} }{\partial \delta P^I\partial \delta P^J}\beta^{\mu J}\right)\delta{P}^I.
\end{split}\end{align}
Clearly, the difference between the lapse and shifts in any two gauges is completely determined by the physical part of the Hamiltonian $\mathcal{H}^{(2)}_{phys}$ and the gauge-invariant coefficient $\lambda_{4}$. As indicated below Eq. \eqref{KHw}, the value of $\lambda_{4}$ can be easily obtained from the algebra of the hypersurface deformations \cite{HOJMAN197688}. 

Furthermore, it is straightforward to deduce how the three-geometries transform under the gauge transformations. Consider the following linear map:
\begin{align}\label{reconM}
\left(\begin{array}{c}\delta \mathcal{H}_\mu\\ \delta C^{\mu}\\ \delta Q_I\\ \delta P^I\end{array}\right) =\mathbf{M}\left(\begin{array}{c}\delta q_{ab}\\ \\  \delta \pi^{ab}\end{array}\right),
\end{align}
where $\mathbf{M}$ is a matrix of the background coefficients computed for the preferred gauge-fixing functions $\delta C^{\mu}$. Then the physical three-surface is obtained from the vanishing of $\delta C^{\mu}$:
\begin{align}
\left(\begin{array}{c}\delta q_{ab}\\ \\ \delta \pi^{ab}\end{array}\right)=\mathbf{M}^{-1}\left(\begin{array}{c}0\\ 0\\ \delta Q_I\\ \delta P^I\end{array}\right).
\end{align}
It turns out that the physical three-surface in any gauge $\delta \tilde{C}^{\mu}=0$ reads
\begin{align}\label{ab3g}
\left(\begin{array}{c}\delta \tilde{q}_{ab}\\ \\ \delta \tilde{\pi}^{ab}\end{array}\right)=\mathbf{M}^{-1}\left(\begin{array}{c}0\\ -\alpha^\mu_{~I}\delta P^I
	 -\beta^{\mu I} \delta Q_{I}\\ \delta Q_I\\ \delta P^I\end{array}\right),
\end{align}
and is a linear function of the coefficients $\alpha^\mu_{~I}$ and $\beta^{\mu I}$.

Let us summarize the obtained procedure for deriving spacetime solutions for arbitrary gauges. First, we set up a {\bf gauge frame} by placing a preferred gauge at the point of origin. Then, the respective full spacetime metric for that particular gauge is computed as explained in Sec. \ref{ISR}. The usual and convenient choice for the point of origin is the spatially flat (or, spatially uniform) gauge. Next, we conveniently construct all other gauges by arbitrary choices of the parameters $\alpha^\mu_{~I}$ and $\beta^{\mu I}$ of a fixed gauge frame. As showed, they completely determine the new full spacetime metric: (i) they determine the lapse function and the shift vector on the three-surfaces through the stability equation \eqref{lapseshift}; and (ii) they define the metric of the three-surfaces via Eq. \eqref{ab3g}.

\subsection{Partial gauge-fixing}
It is sometimes useful to fix a gauge in other ways than by explicitly setting gauge-fixing conditions $\delta C^{\mu}=0$. For instance, we may impose the synchronous gauge, that is, specify the spacetime coordinate system by means of conditions on the lapse and shift functions. More generally, we may replace only some of the gauge-fixing conditions with conditions on the lapse and shift functions. Nevertheless, we find it sufficient to restrict our attention to the case of 4 conditions on the lapse and shift functions. We shall call this method `partial gauge-fixing' to distinguish it from the method used in the last subsection. 
 
Let us first observe an interesting analogy with the well-known gauge theory in electrodynamics (see e.g. \cite{Heitler}, \cite{Weinberg:1995mt}). Indeed, the procedures of gauge-fixing in electrodynamics and cosmological perturbation theory are very similar. In electrodynamics, the Coulomb gauge, $\nabla\vec{A}=0$, is an example of a gauge-fixing condition on the kinematical phase space made of the spatial components of the four-potential and their conjugate momenta $(\vec{A},\vec{\pi})$. On the other hand, the Lorenz gauge, $\partial_\mu A^\mu=0$, is an example of a partial gauge-fixing condition on the temporal component of the four potential $A^{0}$. The latter plays a role of the Lagrange multiplier analogously to the lapse and shift, and multiplies the only constraint of electrodynamics, the Gauss constraint. As we will see below, the partial gauge-fixing in the present theory respects a limited amount of covariance, which is a clear counterpart of the Lorentz-invariance of the Lorenz gauge in electrodynamics.

\begin{figure}[H]

	\begin{center}
		\begin{tikzpicture}
			
			\draw [color=black, ->] (0,0)--(0,3)
			node[label={[xshift=0 cm, yshift=0cm]\scriptsize{$\delta P$}}]{};
					\draw [color=black, ->] (0,0)--(3,0)
			node[label={[xshift=0.4cm, yshift=-0.4cm]\scriptsize{$\delta Q$}}]{};
			
			\filldraw[black] (0,0) circle (1pt) node[label={[xshift=-0.4cm, yshift=-0.7cm]\scriptsize{\textit{Initial gauge}}}]{};
			
				\draw [color=black,-latex] (0,0)--(1.5,2)
			 node[label={[xshift=0.2cm, yshift=-0.1cm]\scriptsize{\textit{New gauge}}}]{};
			
			
			\draw [color=gray] (1.5,2)--(1.5,0)
			 node[label={[xshift=0cm, yshift=-0.6cm]\scriptsize{$\beta$}}]{};
			
			
			\draw [color=gray] (1.5,2)--(0,2)
			node[label={[xshift=-0.2cm, yshift=-0.3cm]\scriptsize{$\alpha$}}]{};
			
		\end{tikzpicture}
	\end{center}
	\caption{{A displacement vector in the space of gauge-fixing conditions determines a new gauge via a shift from the initial gauge at the point of origin. }}
	\label{fig:gauge}
\end{figure}
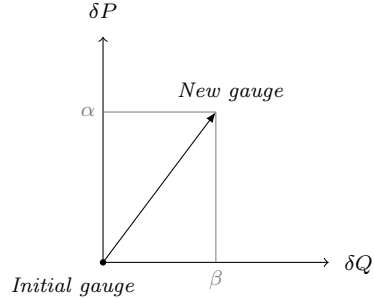

Let us first study the subspace of gauge transformations that preserve the lapse and shift functions, that is, $\frac{\delta\tilde{N}^\mu}{N}\big|_{\delta\tilde{C}^{\mu}=0}- \frac{\delta{N}^\mu}{N}\big|_{\delta{C}^{\mu}=0}=0$. This will determine the residual gauge freedom associated with this method of gauge-fixing. Making use of Eq. \eqref{lapseshift} we find the ambiguity in the choice of the respective gauge-fixing conditions, here expressed in terms of  $\alpha^\mu_{~I}$ and $\beta^{\mu I}$ satisfying the following dynamical equations (for each $\vec{k}$):
%
%

\begin{align}\label{kernel}\begin{split}
	\dot{\alpha}^\mu_{~I} &=-
	\beta^{\mu J}
	\frac{\partial^2\mathcal{H}^{(2)}_{phys} }{\partial \delta P^J\partial \delta P^I}
	+\alpha^\mu_{~J}
	\frac{\partial^2\mathcal{H}^{(2)}_{phys} }{\partial \delta Q_J\partial \delta P^I}-\lambda_{4\nu}^\mu\alpha^\nu_{~I},\\
	\dot{\beta}^{\mu I} &=
	-\beta^{\mu J}
	\frac{\partial^2\mathcal{H}^{(2)}_{phys} }{\partial \delta P^J\partial \delta Q_I}
	+\alpha^\mu_{~J}
	\frac{\partial^2\mathcal{H}^{(2)}_{phys} }{\partial \delta Q_J\partial \delta Q_I}-\lambda_{4\nu}^\mu\beta^{\nu I},
	\end{split}
\end{align}
where the second-order partial derivatives yield the background coefficients of the physical Hamiltonian. Note that the solution does not depend on the particular choice of the lapse and the shifts. Once $\alpha^\mu_{~I}$ and $\beta^{\mu I}$ are set at an initial time for all $I={1,2}$ and $\mu={0,1,2,3}$, a unique solution $t\mapsto (\alpha^\mu_{~I}(t),\beta^{\mu I}(t))$ exists. Hence, at the initial time $t_0$ we have complete freedom in defining the gauge-fixing functions,
\begin{align}
\delta \tilde{C}^{\mu}(t_0)\approx \delta C^\mu(t_0) +\alpha^\mu_{~I}(t_0)\delta P^I+\beta^{\mu I} (t_0)\delta Q_{I},
\end{align}
where $\delta C^\mu(t_0)$ lies at the point of origin of the gauge frame and $\delta \tilde{C}^{\mu}(t_0)$ are arbitrary gauge-fixing functions. Once the choice is made, Eq. \eqref{kernel} determines the gauge-fixing functions at all other times. There is a very clear spacetime picture associated with this ambiguity (see Fig. \ref{fig:gauge}): once $\delta \tilde{C}^{\mu}(t_0)$ are chosen, the initial three-surface with coordinates on it is fixed. If the initial values of the gauge-invariant variables $(\delta Q_{I}(t_0), \delta P^I(t_0))$ are known then the initial three-surface may be reconstructed explicitly in terms of the ADM perturbation variables.

Furthermore, the evolution of the three-surface with its coordinates is uniquely determined via the evolution of $\delta \tilde{C}^{\mu}(t)$ and the independent evolution of the gauge-invariant variables $(\delta Q_{I}(t), \delta P^I(t))$. Hence the full spacetime geometry is reconstructed. Note the very important feature that the spacetime coordinate system is introduced in a way that is independent of the evolution of the gauge-invariant variables $(\delta Q_{I}(t), \delta P^I(t))$.

Now let us consider the case in which the LHS of Eq. \eqref{lapseshift} is non-vanishing. Then $\frac{\delta\tilde{N}^\mu}{N}\big|_{\delta\tilde{C}^{\mu}=0}- \frac{\delta{N}^\mu}{N}\big|_{\delta{C}^{\mu}=0}$ is an arbitrary linear combination of Dirac observables. In this case Eq. \eqref{lapseshift} implies
\begin{align}\begin{split}\label{alphabetadot}
\dot{\alpha}^\mu_{~I}
	=
	-
	\beta^{\mu J}
	\frac{\partial^2\mathcal{H}^{(2)}_{phys} }{\partial \delta P^J\partial \delta P^I}
	+\alpha^\mu_{~J}
	\frac{\partial^2\mathcal{H}^{(2)}_{phys} }{\partial \delta Q_J\partial \delta P^I}-\lambda_{4\nu}^\mu\alpha^\nu_{~I}
-\frac{\partial}{\partial \delta P^{I}}\left(\frac{\delta \tilde{N}^\mu-\delta{N}^\mu}{N}\right),\\
\dot{\beta}^{\mu I}
	=
	-\beta^{\mu J}
	\frac{\partial^2\mathcal{H}^{(2)}_{phys} }{\partial \delta P^J\partial \delta Q_I}
	+\alpha^\mu_{~J}
	\frac{\partial^2\mathcal{H}^{(2)}_{phys} }{\partial \delta Q_J\partial \delta Q_I}-\lambda_{4\nu}^\mu\beta^{\nu I}
-\frac{\partial}{\partial \delta Q_{I}}\left(\frac{\delta \tilde{N}^\mu-\delta{N}^\mu}{N}\right).\end{split}
\end{align}

A unique solution is obtained by assuming initially $\alpha^\mu_{~I}(t_0)={\beta}^{\mu I}(t_0)=0$. The complete space of solutions is then constructed by combining it with the solutions of Eq. (\ref{kernel}).

\section{Perturbed Kasner universe}\label{sectionkasner}
Below we illustrate the obtained results by considering the Hamiltonian theory of cosmological perturbations in the Kasner universe. The latter is the vacuum limit of the Bianchi Type I model with a scalar field, which was recently considered in \cite{boldrin2021dirac}. The reader can consult the mentioned article for more details.

\subsection{Background model}
The metric of the Kasner universe reads:
\begin{align}
	\ud s^2=-\ud t^2+\sum_ia_i^2(\ud x^i)^2,~~~a=(a_1a_2a_3)^{\frac{1}{3}},
\end{align}
where we assume the coordinates $(x^1,x^2,x^3)\in [0,1)^3$. The $x^i$-axes are the principal axes in which the extrinsic curvature of the spatial leaves is diagonal, nevertheless the expansion can occur at different rates along different axes. The three-momentum is diagonal too, and the Hamiltonian constraint reads:
\begin{align}\label{H0}
\mathcal{H}_{Kas}=a^{-3}\bigg(\frac{1}{2}\sum_i(a_i^2p^i)^2-\sum_{i> j}a_i^2p^ia_j^2p^j\bigg),
\end{align}
where $\{a_i^2,p^j\}=\delta_{i}^{~j}$. The Hamilton equations read:
\begin{align}\label{backeom}\begin{split}
\dot{p}^{i}=-\frac{1}{a^3}\bigg((a_ip^i)^2-\sum_{j\neq i}p^{i}a_j^2p^j\bigg),
\quad
\dot{a}_{i}=\frac{a_{i}}{2a^{3}}\bigg(a_{i}^2p^{i}-\sum_{j\neq i}a_j^2p^j\bigg),\end{split}
\end{align}
where the initial data $(a_i^2(t_0), p^i(t_0))$ are assumed to lie in the constraint surface, $\mathcal{H}_{Kas}=0$.

For studying the three-metric and three-momentum perturbations propagating along a wavevector $\vec{k}=(k^1,k^2,k^3)$ it is more convenient to switch from the coordinate basis to a basis of three orthonormal vectors $(\hat{k},\hat{v},\hat{w})$. The three vectors are normalized with the conformal metric, $\gamma_{ij}=\frac{a_i^2}{a^2}\delta_{ij}$, and so $\hat{k}^i=k^i/\sqrt{k^ik^j\gamma_{ij}}$. We use the triad $(\hat{k},\hat{v},\hat{w})$ to replace the components of the background-level three-momentum $(p^1,p^2,p^3)$ with a nondiagonal tensor $P$ whose components read $P_{nm}=a^2\sum_ip^i\hat{n}_i\hat{m}_i$,  where $\hat{n},\hat{m}\in (\hat{k},\hat{v},\hat{w})$. One may show that $P$ is related to the shear tensor as follows $\sigma_{nm}=a^{-2}P_{{n}{m}}-\frac{a^{-2}}{3}TrP\cdot\delta_{{n}{m}}$, where the shear is defined as $\sigma_{{n}{m}}=\sigma_{ij}\hat{n}^i\hat{m}^j$ and $\sigma_{ij}=\frac{1}{2}\frac{d}{d\eta}\big(\frac{a_i^2}{a^2}\big)\delta_{ij}$. The Hamiltonian constraint $\mathcal{H}_{Kas}$ now reads:
\begin{align}
	\mathcal{H}_{Kas}=a^{-3}
	\left((TrP^2)-\frac{1}{2}(Tr P)^2\right).
\end{align}

Note that the choice of $(\hat{v},\hat{w})$ is free,  nevertheless, it is convenient to impose:
\begin{align}\label{fermilaw}
\frac{\ud \hat{v}^j}{\ud\eta}=-\sigma_{{v}{v}}\hat{v}^j-\sigma_{{v}{w}}\hat{w}^j,~~\frac{\ud \hat{w}^j}{\ud\eta}=-\sigma_{{w}{w}}\hat{w}^j-\sigma_{{w}{v}}\hat{v}^j,
\end{align} 
where $\eta$ denotes conformal time. For more details, see \cite{boldrin2021dirac} or \cite{Uzan}.
 
\subsection{Perturbations} \label{pert}
We incorporate into the Kasner universe purely inhomogeneous three-metric and three-momentum perturbations, $\delta q_{ij}$ and $\delta\pi^{ij}$. The lapse function and the shift vector are also perturbed with the purely inhomogeneous $\delta N$ and $\delta N^i$. We switch to the momentum representation of the perturbations: $\delta \check{q}_{ij}(\underline{k})=\int \delta{q}_{ij}(\vec{x})e^{-ik_ix^i}\ud^3x$, $\delta \check{\pi}^{ij}(\underline{k})=\int \delta{\pi}^{ij}(\vec{x})e^{-ik_ix^i}\ud^3x$, $\delta \check{N}(\underline{k})=\int \delta{N}(\vec{x})e^{-ik_ix^i}\ud^3x$ and $\delta \check{N}^{i}(\underline{k})=\int \delta{N}^{i}(\vec{x})e^{-ik_ix^i}\ud^3x$. {The reality condition for the perturbations reads: $\delta \check{q}_{ij}(\underline{k})=\delta \bar{\check{q}}_{ij}(-\underline{k})$ and $\delta \check{\pi}^{ij}(\underline{k})=\delta \bar{\check{\pi}}^{ij}(-\underline{k})$.} Furthermore, we decompose all the (spatial) tensors in a basis given by the triad $(\hat{k},\hat{v},\hat{w})$. In particular, any symmetric 2-rank tensor is decomposed into the following homogeneous tensor fields: $A_{ij}^1=\gamma_{ij}$, $A_{ij}^2=\hat{k}_i\hat{k}_j-\frac{1}{3}\gamma_{ij}$, $A_{ij}^3=\frac{1}{\sqrt{2}}\Big(\hat{k}_i\hat{v}_j+\hat{v}_i\hat{k}_j\Big)$, $A_{ij}^4=\frac{1}{\sqrt{2}}\Big(\hat{k}_i\hat{w}_j+\hat{w}_i\hat{k}_j\Big)$, $A_{ij}^5=\frac{1}{\sqrt{2}}\Big(\hat{v}_i\hat{w}_j+\hat{w}_i\hat{v}_j\Big)$, $A_{ij}^6=\frac{1}{\sqrt{2}}\Big(\hat{v}_i\hat{v}_j-\hat{w}_i\hat{w}_j\Big)$. The dual tensors $A_n^{ij}$, $n=1,\dots, 6$ are defined in such a way that $A_n^{ij}A^m_{ij}=\delta_{n}^m$. Finally, our definitions of the perturbation variables read: $\delta {q}_{n}(\underline{k})=\delta \check{q}_{ij}A_n^{ij}$, $\delta {\pi}^{n}(\underline{k})=\delta \check{\pi}^{ij}A^n_{ij}$, $\delta {N}^{k}(\underline{k})=\delta \check{N}_i\hat{k}^i$, $\delta N^{v}(\underline{k})=\delta \check{N}_i\hat{v}^i$ and $\delta N^{w}(\underline{k})=\delta \check{N}_i\hat{w}^i$. 

The dynamical variables satisfy the following commutation relations:
\begin{align}
\{\delta {q}_{n}(\underline{k}),\delta{\pi}^{m}(-\underline{k}')\}=\delta^m_n\delta_{\underline{k},\underline{k}'}.
\end{align}
The Hamiltonian \eqref{Htot} specified to the Kasner universe and expressed with the help of the triad $(\hat{k},\hat{v},\hat{w})$ reads
\begin{align}\label{Ktot}
		\mathbb{H}=\int \left( N\mathcal{H}_{Kas}+N\mathcal{H}^{(2)}+\delta N\delta \mathcal{H}_0+\delta N^k\delta \mathcal{H}_k+\delta N^v\delta \mathcal{H}_v+\delta N^w\delta \mathcal{H}_w\right)\ud^3x,
\end{align}
where $\mathcal{H}^{(2)}$, $\delta \mathcal{H}_0$, $\delta \mathcal{H}_k$, $\delta \mathcal{H}_v$ and $\delta \mathcal{H}_w$ are given in Appendix \ref{kasner}.

\subsection{Spatially flat gauge}\label{sfs}
The spatially flat gauge is given by the following set of gauge-fixing functions:
\begin{align}\label{sfg}\delta c_1=\delta q_1,~\delta c_2=\delta q_2,~\delta c_3=\delta q_3,~\delta c_4=\delta q_4.\end{align}
We use both the constraints $\delta \mathcal{H}_0= 0$, $\delta \mathcal{H}_k= 0$, $\delta \mathcal{H}_v= 0$, $\delta \mathcal{H}_w= 0$ and the gauge-fixing conditions \eqref{sfg} to replace $\delta \pi^1$, $\delta \pi^2$, $\delta \pi^3$ and $\delta \pi^4$ by linear combinations of  $\delta q_5$, $\delta q_6$, $\delta \pi^5$ and $\delta \pi^6$. One advantage of this choice is that the Dirac bracket is immediately obtained:
\[\{\cdot,\cdot\}_D=\sum_{\underline{k}}\sum_{i=5,6}\frac{\partial~\cdot}{\partial\delta {q}_{i}(\underline{k})}\frac{\partial~\cdot}{\partial\delta {\pi}^{i}(-\underline{k})}-\frac{\partial~\cdot}{\partial\delta {\pi}^{i}(-\underline{k})}\frac{\partial~\cdot}{\partial\delta {q}_{i}(\underline{k})}.\]
The second-order Hamiltonian \eqref{Ktot} becomes a quadratic function of the perturbations $\delta q_5$, $\delta q_6$, $\delta \pi^5$ and $\delta \pi^6$ only. Furthermore, we find it convenient to use the rescaled variables $\delta \tilde{q}_{5}=\frac{1}{\sqrt{2}a}\delta q_{5}$, $\delta \tilde{q}_{5}=\frac{1}{\sqrt{2}a}\delta q_{5}$, $\delta \tilde{\pi}^{5}=\sqrt{2}a\delta \pi^{5}$ and $\delta \tilde{\pi}^{6}=\sqrt{2}a\delta \pi^{6}$. Finally, the perturbation Hamiltonian is found to read (see \cite{boldrin2021dirac} for details):
\begin{align}\label{HK2}\begin{split}
\mathcal{H}^{(2)}_{red}=
\frac{N}{2a}\bigg[\delta\tilde{\pi}_5^2+\delta\tilde{\pi}_6^2+(k^2+U_1)\delta \tilde{q}_5^2+(k^2+U_2)\delta\tilde{q}_6^2+C_{12}\delta \tilde{q}_5\delta \tilde{q}_6\bigg],
\end{split}
\end{align}
where the coefficients can be found in Appendix \ref{coeff}. 

The above Hamiltonian can be used to formulate the gauge-invariant dynamics of perturbations. To this end we find 4 independent Dirac observables, denoted by $\delta Q_1$, $\delta Q_2$, $\delta P^1$ and $\delta P^2$ as linear combinations of the ADM perturbation variables that Poisson-commute with the constraint functions $\delta \mathcal{H}_0$, $\delta \mathcal{H}_k$, $\delta \mathcal{H}_v$, $\delta \mathcal{H}_w$ and the gauge-fixing functions $\delta c_1$, $\delta c_2$, $\delta c_3$, $\delta c_4$. They read
\begin{align}\label{QPdirac}
	\begin{aligned}
\delta Q_1&=
{\frac{1}{\sqrt{2} a}\delta q_5}
+{\frac{2P_{vw}}{a P_{kk}}}
{(\delta q_1-\frac{1}{3}\delta q_2)},
\\\delta Q_2&=
 {\frac{1}{\sqrt{2} a}\delta q_6}
 +{\frac{P_{vv}-P_{ww}}{a P_{kk}}}
 {(\delta q_1-\frac{1}{3}\delta q_2)},
  \\
\delta P^1&=
{\sqrt{2} a\delta\pi_5+\frac{\frac{5}{6} (Tr P)- P_{kk}}{\sqrt{2} a^3}\delta q_5}
-{\frac{2P_{vw}}{\sqrt{2} a^3 P_{kk}}}
{\left(\frac{P_{vv}-P_{ww}}{2}\delta q_6+P_{vw}\delta q_5\right)}
\\
&+{\mathcal{F}(P_{vw},P_{kv} P_{kw})}
{(\delta q_1-\frac{1}{3}\delta q_2)}
-{\frac{3P_{vw}}{a^3P_{kk}}}
{P_{kk}\delta q_1}
+\frac{\sqrt{2}}{a^3}\left(P_{kw} \delta q_3+P_{kv}\delta q_4\right),
\\
 \delta P^2&=
 {\sqrt{2} a\delta\pi_6+\frac{\frac{5}{6} (Tr P)- P_{kk}}{\sqrt{2} a^3}\delta q_6}
  -\frac{P_{vv}-P_{ww}}{\sqrt{2} a^3 P_{kk}}\left(\frac{P_{vv}-P_{ww}}{2}\delta q_6+P_{vw}\delta q_5\right)
 \\
 &+{\mathcal{F}\left(\frac{P_{vv} - P_{ww}}{2},\frac{P_{kv}^2-P_{kw}^2}{2}\right)}(\delta q_1-\frac{1}{3}\delta q_2)
 -{\frac{3(P_{vv}-P_{ww})}{2a^3}}\delta q_1\\
 &+\frac{\sqrt{2}}{a^3}\left(P_{kv} \delta q_3-P_{kw}\delta q_4\right),
\end{aligned}
\end{align}
where $\mathcal{F}(X,Y)=\frac{4}{a^3 P_{kk}}Y-\frac{4 \left(P_{vw}^2+ (\frac{P_{vv} - P_{ww}}{2})^2\right) -2P_{kk}(P_{kk}+\frac{Tr P}{3})}{a^3 P_{kk}^2}X$. It is easy to show that these Dirac observables transform as spin-$2$ fields under rotations, that is, $\delta Q_1$ and $\delta Q_2$ are two polarization modes of the gravitational wave. Note that in the gauge-fixing surface \eqref{sfg} these Dirac observables are numerically equal to the rescaled ADM perturbation variables, that is,
\begin{align*}
 \delta{Q}_1\big|_{SF}=\delta\tilde{q}_5,\quad \delta{Q}_2\big|_{SF}=\delta\tilde{q}_6, \quad \delta{P}^1\big|_{SF}=\delta\tilde{\pi}_5,\quad \delta{P}^2\big|_{SF}=\delta\tilde{\pi}_6,
 \end{align*}
 where `SF' denotes the spatially flat gauge. Therefore, in terms of the Dirac observables the Hamiltonian \eqref{HK2} becomes
\begin{align}\begin{split}\label{H2dirac}
		\mathcal{H}^{(2)}_{phys}=
		\frac{N}{2a}\bigg[
		(\delta{P}^{1})^2+(\delta{P}^2)^2+(k^2+U_1)\delta{Q}_1^2+(k^2+U_2)\delta{Q}_2^2+C_{12}\delta{Q}_1\delta{Q}_2\bigg]. 
	\end{split}
\end{align}
It generates the gauge-invariant dynamics of the system as discussed in Sec. \ref{gid}.

In order to reconstruct the actual spacetime we apply the formula \eqref{coneq} and find
\begin{align}\label{NB}\begin{split}
\frac{\delta N}{N}&=
-\frac{P_{vw} }{ a P_{kk}}\delta Q_1
-\frac{P_{vv}-P_{ww}}{2 a P_{kk}}\delta Q_2,\\
	\frac{1}{i}\frac{\delta N^k}{N}&=
\bigg(
\frac{ 2P_{vw}}{3 a^4} 
-\frac{ 2 P_{kv} P_{kw} }{a^4 P_{kk}}
+\frac{ 2P_{vw}^3}{a^4P_{kk}^2}+\frac{ P_{vw}(P_{vv}-P_{ww})^2}{2a^4P_{kk}^2}-\frac{ 5P_{vw}(P_{vv}+P_{ww})}{6a^4P_{kk}}\bigg)\delta Q_1\\
&+\bigg(
\frac{ P_{kw}^2}{a^4 P_{kk}} 
-\frac{ P_{kv}^2 }{a^4 P_{kk}}
+\frac{ P_{vv}-P_{ww} }{3 a^4}+\frac{ 5(P_{ww}^2 -P_{vv}^2)}{12 a^4P_{kk}}+\frac{(P_{vv}-P_{ww})P_{vw}^2}{a^4P_{kk}^2}+\frac{(P_{vv}-P_{ww})^3}{4a^4P_{kk}^2}
\bigg)\delta Q_2
\\&+\frac{ P_{vw}}{a^2 P_{kk}}\delta P^1
+\frac{ P_{vv}-P_{ww}}{2a^2 P_{kk}}\delta P^2
,\\
\frac{1}{i}\frac{\delta N^v}{N}&=
\bigg(
\frac{ 2 P_{kv} P_{vw} }{a^4 P_{kk}}
+\frac{ 2 P_{kw}}{a^4}
\bigg)\delta Q_1
+
\bigg(
\frac{ P_{kv} (P_{vv}-P_{ww}) }{a^4 P_{kk}}
+\frac{ 2 P_{kv} }{a^4}
\bigg)\delta Q_2
,\\
\frac{1}{i}\frac{\delta N^w}{N}&=
\bigg(
\frac{2 P_{kw} P_{vw}}{a^4 P_{kk}}
+\frac{ 2 P_{kv}}{a^4}
\bigg)\delta Q_1
+
\bigg(
\frac{ P_{kw} (P_{vv}- P_{ww})}{a^4 P_{kk}}
-\frac{2 P_{kw} }{a^4}
\bigg)\delta Q_2.
\end{split}
\end{align}
{(Note that the imaginary units in front of $\delta N^k$, $\delta N^v$, $\delta N^w$ cancel out in Eq. \eqref{Ktot} when multiplied by the respective constraints \eqref{linD}).}
\subsection{Gauge transformations}
In what follows we consider two other gauges for the illustration of the method.

\subsubsection{Scalar gravity gauge}
Let us first consider a gauge in which one of the tensor modes of the metric perturbation vanishes, say $\delta q_5=0$. Moreover,  we assume $\delta q_1=\delta q_3=\delta q_4=0$. Note that in this gauge one polarization mode of the gravitational wave, $\delta Q_1=-{\frac{2P_{vw}}{3aP_{kk}}}{\delta q_2}$, is carried by a scalar metric perturbation. Thus, we shall call it the scalar gravity gauge (SG). To make use of the gauge frame based on the spatially flat gauge (SF) we cast both sets of gauge-fixing functions into canonical form (see Eq. \eqref{can}):
{\small \begin{align}\begin{split}
\delta C_{SF}^\mu=\Bigg(&
	\frac{a (3 \delta q_1-\delta q_2)}{3 P_{kk}}
	,
	\frac{i \left[(P_{vv}+P_{ww})(3 \delta q_1-\delta q_2) -P_{kk}(6\delta q_1+\delta q_2)\right]}{6a^2 P_{kk}}
	,\nonumber\\
	&
		-\frac{i \left(12 P_{kv}\delta q_1-4 P_{kv}\delta q_2+3 \sqrt{2} P_{kk}\delta q_3\right)}{6a^2 P_{kk}}
			,
		-\frac{i \left(12 P_{kw}\delta q_1-4 P_{kw}\delta q_2+3 \sqrt{2} P_{kk}\delta q_4\right)}{6a^2 P_{kk}}
	\Bigg),\\
		\delta C_{SG}^\mu=\Bigg(&
		-\frac{a\delta q_5}{2 \sqrt{2}P_{vw}}
		,
		-\frac{i \left(12 P_{vw}\delta q_1+\sqrt{2}(TrP)\delta q_5\right)}{8a^2 P_{vw}}
		,
		-\frac{i \left( P_{vw}\delta q_3- P_{kv}\delta q_5\right)}{\sqrt{2}a^2 P_{vw}}
		,
	-\frac{i \left( P_{vw}\delta q_4- P_{kw}\delta q_5\right)}{\sqrt{2}a^2 P_{vw}}
		\Bigg),
\end{split}\end{align}}
and find the difference in terms of $\alpha$ and $\beta$ to read:
\begin{align}\label{betaSGFS}\begin{split}
\alpha^\mu_{~I}=0,~~
	\beta^{\mu I}=
\begin{pmatrix}
-\frac{a^2}{2 P_{vw}}&0\\
\frac{-i (TrP)}{4 a P_{vw}}&0\\
\frac{i P_{kv}}{a P_{vw}}&0\\
\frac{i P_{kw}}{a P_{vw}}&0
\end{pmatrix}.
\end{split}\end{align}
Furthermore, we find $\lambda^\mu_{4\nu}$ to read
\begin{align}
\lambda_4
=
\begin{pmatrix}
0&0&0&0\\
i a^{-2}k&0&0&0\\
0&-\frac{2P_{kv}}{a^3}&\frac{P_{kk}-P_{vv}}{a^3}&-\frac{P_{vw}}{a^3}\\
0&-\frac{2P_{kw}}{a^3}&-\frac{P_{vw}}{a^3}&\frac{P_{kk}-P_{ww}}{a^3}
\end{pmatrix}.
\end{align}
Hence, the lapse and shift transform as follows (see Eq. \eqref{lapseshift}):
\begin{align}\begin{split}
		\frac{\delta{N}_{SG}^\mu}{N}\bigg|_{\delta\tilde{C}_{SG}^{\mu}=0}- \frac{\delta{N}_{SF}^\mu}{N}\bigg|_{\delta{C}_{SF}^{\mu}=0}
		&\approx
		\left(
		\lambda_{4\nu}^\mu\beta^{\nu 1}
		+\dot{\beta}^{\mu 1}
		\right)\delta{Q}_1
		+
\frac{1}{2a}\beta^{\mu 1}\delta{P}^1.
\end{split}\end{align}
To reconstruct the three-surfaces we apply the formula \eqref{ab3g} with the matrix $\mathbf{M}$ which maps the ADM pertrubation variables into the constraint functions \eqref{linS}, \eqref{linD}, the gauge-fixing functions \eqref{sfg} and the Dirac observables \eqref{QPdirac}.

\subsubsection{Synchronous gauge}
The synchronous gauge is given by partial gauge-fixing, $\delta N^\mu=0$.
The gauge-fixing conditions for synchronous gauge are obtained in terms of $\alpha^\mu_{~I}$ and $\beta^{\mu I}$ as solutions to Eqs \eqref{alphabetadot}. In the gauge frame originated on the SF gauge, we find: 
\begin{align}\begin{split}
		\dot{\alpha}^\mu_{~I}
	&	=
		-\frac{1}{2a}
		\beta^{\mu I}
		+\frac{\partial}{\partial \delta P^{I}}\left(\frac{\delta {N}_{SF}^\mu}{N}\right),\\
		\dot{\beta}^{\mu I}
	&	=
		U_I\alpha^\mu_{~I}+C_{IJ}\alpha^\mu_{~J}
		-\lambda_{4\nu}^\mu\beta^{\nu I}
		+\frac{\partial}{\partial \delta Q_{I}}\left(\frac{\delta{N}_{SF}^\mu}{N}\right),\end{split}
\end{align}
where $\frac{\delta{N}_{SF}^\mu}{N}$ are given by Eqs \eqref{NB}. The choice of the initial data ${\alpha}^\mu_{~I}(t_0)$, ${\beta}^{\mu I}(t_0)$ determines unambiguously the synchronous gauge-fixing conditions. Then the three-surfaces can be reconstructed with the use of matrix $\mathbf{M}$ as in the previous example.

\section{Conclusions}\label{conclusion}
The purpose of this work was to develop a complete Hamiltonian approach to CPT. The basic property of our approach is the separation of the gauge-independent dynamics of perturbations from the problem of gauge-fixing and spacetime reconstruction. We use the Dirac procedure for constrained systems to derive the dynamics of gauge-dependent perturbations and to rewrite it in terms of gauge-independent quantities, the Dirac observables. A key element of our approach is the reconstruction of spacetime based on  gauge-fixing conditions. The usual approaches, e.g.  \cite{Mukhanov:1990me,Malik:2008im,sasaki-cpt}, suffer from the lack of methodological choice of gauge-fixing conditions, the choice of gauge is often more like a guess rather than an actual choice made within a well-defined and complete set of possibilities\footnote{ Some commonly used gauges, their validity and, in some cases, their residual freedom are studied in the mentioned references. However, no general method for defining a valid gauge and its residual freedom is provided. In particular, the mentioned references lack a clear exposition of the connection between the residual freedom, the lapse and the shift, and the choice of the initial three-surface.}. To overcome this problem, we introduced the Kucha\v r decomposition for the ADM perturbation phase space. The space of all the possible gauge-fixing conditions and the gauge transformations induced by the linear diffeomorphisms of three-surfaces are made explicit via this decomposition. Moreover, it makes the transformations of the lapse and shift manifestly dependent on purely gauge-independent terms of the full Hamiltonian. This simplifies the problem of spacetime reconstruction and provides a tool for studying partial gauge-fixing.

To illustrate our approach we consider the perturbed Kasner universe. We established a gauge frame in the space of all gauges for this model: we chose the spatially flat gauge as the point of origin and the Dirac observables \eqref{QPdirac} as basic displacements. We first discussed a gauge-fixing condition that kills one of the tensor modes of the three-metric perturbation. In this gauge one polarization mode of the gravitational wave is carried by a pure scalar metric perturbation. We expressed this gauge in the gauge frame in order to reconstruct the full spacetime metric. Our next example was the synchronous gauge that is given via partial gauge-fixing conditions on the lapse and shift and for which gauge-fixing conditions remain underdetermined. This is an analog of the Lorenz gauge in electrodynamics. We showed how to use the gauge frame to determine gauge-fixing conditions and how to use the gauge-fixing conditions to reconstruct the spacetime.

The possible applications of the presented Hamiltonian formalism include addressing key conceptual problems in quantum cosmology such as the time problem, the semiclassical spacetime reconstruction, or the relation between the kinematical and reduced phase space quantization. The full clarification of the Hamiltonian formalism and its structure is essential for these and similar tasks.

{ We note that cosmological perturbation theory owes its simplicity to the abelianization of the algebra of hypersurface deformations in linear approximation. The abelianization of the constraints' algebra is specific to perturbation theory, although it could also occur in the context of other-type truncations to the full canonical formalism. Therefore, we do not expect the abelianization of the constraints' algebra to occur when expanding the constraints beyond linear order. However, the abelianization of constraints in linear perturbation theory can be extended to higher orders in the successive order-by-order expansion. At each successive order one introduces new perturbation variables and derives linear equations of motion for them while assuming a fixed solution at lower orders. The linear equations of motion can be cast to the form of Hamilton equations generated by a quadratic (in the new variables) Hamiltonian with the use of a symplectic form independent of the lower-order symplectic forms. Therefore, the structure of the Hamiltonian formalism is expected to reproduce at each order (except for the zero order). At each order linearized gauge transformations are distinct as they act on separate perturbation variables. Order by order they add up to form a non-linear, `curved’ space of gauge-fixing conditions. This theory, however, by assumption cannot be extended to perturbations large enough to change the topology of gauge-fixing conditions. Therefore, we do not foresee any obstructions in gauge-fixing in higher-order perturbation theory either. Nevertheless, the details of this higher-order theory and of the role the Kucha\v r parametrization could play in it can be found only via a careful derivation. We postpone a more detailed analysis of this approach to a future paper.
}

\begin{acknowledgments}
The authors acknowledge the support from the National Science Centre (NCN, Poland) under the research grant 2018/30/E/ST2/00370.
\end{acknowledgments}

\appendix

\section{Constraints}\label{kasner}

Note the identity $\delta^{i}_{~j}=\hat{k}^i \hat{k}_j+\hat{v}^i \hat{v}_j+\hat{w}^i \hat{w}_j$. It is convenient to introduce  $\delta \mathcal{H}_k=\delta\mathcal{H}^{~i}_g\hat{k}_i$, $\delta \mathcal{H}_v=\delta\mathcal{H}^{~i}_g\hat{v}_i$, $\delta \mathcal{H}_w=\delta\mathcal{H}^{~i}_g\hat{w}_i$, $\delta N^{k}=\delta N_i\hat{k}^i$, $\delta N^{v}=\delta N_i\hat{v}^i$ and $\delta N^{w}=\delta N_i\hat{w}^i$. The first-order scalar constraint reads:
{\small\begin{align}\label{linS}
		\begin{aligned}
			\delta\mathcal{H}_{0}
			&=
			-\frac{1}{3}a^{-1}(Tr P)\delta\pi^1
			+a^{-1}[3P_{kk}
			-(Tr P)]\delta\pi^2
			+a^{-1}2\sqrt{2}P_{kv}\delta\pi^3
			+a^{-1}2\sqrt{2}P_{kw}\delta\pi^4
			\\
			&\quad
			+a^{-1}2\sqrt{2}P_{vw}\delta\pi^5
			+a^{-1}\sqrt{2}(P_{vv}
			-P_{ww})\delta\pi^6
			+\frac{1}{2}a^{-5}[(Tr P^2)
			-\frac{1}{2}(Tr P)^2]\delta q_1
			\\
			&\quad
			+\frac{1}{3}a^{-5}\{-2(Tr P^2)
			+6(P_{kk}^2+P_{kv}^2+P_{kw}^2)
			-(Tr P)\left[3P_{kk}-(Tr P)\right]\}\delta q_2
			\\
			&\quad
			+\sqrt{2}a^{-5}[2P_{kw} P_{vw}+P_{kv} (P_{vv}-P_{ww})+P_{kk} P_{kv}]\delta q_3
			\\
			&\quad
			+\sqrt{2}a^{-5}[2P_{kv}P_{vw}-P_{kw}(P_{vv}-P_{ww})+P_{kk}P_{kw}]\delta q_4
			\\
			&\quad
			+\sqrt{2}a^{-5}[2P_{kv}P_{kw}-2P_{kk}P_{vw}
			+(Tr P)P_{vw}]\delta q_5
			\\
			&\quad
			+\frac{1}{\sqrt{2}}a^{-5}[2(P_{kv}^2-P_{kw}^2)
			-2P_{kk}(P_{vv}-P_{ww})
			+(Tr P)(P_{vv}-P_{ww})
			]\delta q_6
			\\
			&\quad
			-2a^{-1}k^2(\delta q_1
			-\frac{1}{3}\delta q_2),
		\end{aligned}
\end{align}}
and the first-order vector constraints read:
{	\small{\begin{align}\label{linD}
			\begin{aligned}
				\delta\mathcal{H}_k=&
				-2ia^2
				\bigg[\frac{1}{3}\delta\pi^1
				+\delta\pi^2+a^{-4}(P_{kk}
				-\frac{1}{2}(Tr P))\delta q_1
				+\frac{a^{-4}}{6}(P_{kk}
				+(Tr P))\delta q_2
				\\
				&\qquad
				-\frac{a^{-4}}{\sqrt{2}}P_{vw}\delta q_5
				-\frac{a^{-4}}{\sqrt{2}}\frac{P_{vv}-P_{ww}}{2}\delta q_6\bigg],
				\\
				\delta\mathcal{H}_v=&-2ia^2
				\left[\frac{1}{\sqrt{2}}\delta\pi^3
				+a^{-4}P_{kv}
				\left(
				\delta q_1
				-\frac{1}{3}\delta q_2
				\right)
				+\frac{a^{-4}}{\sqrt{2}}P_{kk}\delta q_3
				+\frac{a^{-4}}{\sqrt{2}}P_{kw}\delta q_5
				+\frac{a^{-4}}{\sqrt{2}}P_{kv}\delta q_6
				\right],
				\\
				\delta\mathcal{H}_w=&-2ia^2
				\left[\frac{1}{\sqrt{2}}\delta\pi^4
				+a^{-4}P_{kw}
				\left(
				\delta q_1
				-\frac{1}{3}\delta q_2
				\right)
				+\frac{a^{-4}}{\sqrt{2}}P_{kk}\delta q_4
				+\frac{a^{-4}}{\sqrt{2}}P_{kv}\delta q_5
				-\frac{a^{-4}}{\sqrt{2}}P_{kw}\delta q_6
				\right].
			\end{aligned}
\end{align}}}
The second-order non-vanishing part of the Hamiltonian \eqref{Ktot} reads
{\small
\begin{align}
	&\mathcal{H}_{0}^{(2)}=
	-\frac{a \delta \pi_1^2}{6}
	+\frac{3 a \delta\pi_2^2}{2}
	+a \delta \pi_3^2
	+a \delta \pi_4^2
	+a \delta \pi_5^2
	+a \delta \pi_6^2
	-
	\delta q_1^2\bigg(
	\frac{k^2}{2 a^3}
	+\frac{47 (Tr P)^2}{16a^7}
	+\frac{(Tr P^2)}{8 a^7}\bigg) 
			\nonumber
	\\
	&
	+\delta q_2^2
	\bigg(-\frac{k^2}{18 a^3}
	-\frac{P_{kk}^2}{6 a^7}
	-\frac{2 P_{kv}^2}{3a^7}
	-\frac{2 P_{kw}^2}{3 a^7}
	+
	\frac{P_{kk} (Tr P)}{3	a^7}
	-\frac{5(Tr P)^2}{36 a^7}
	+\frac{5 (Tr P^2)}{18 a^7}
	\bigg)
	\nonumber
	\\
	&
	+\delta q_3^2
	\bigg(
	-\frac{(Tr P)^2}{8a^7}
	+\frac{P_{kk} P_{vv}}{a^7}
	+\frac{(Tr P^2)}{4a^7}
	\bigg)
	+\delta q_4^2 
	\bigg(
	-\frac{(Tr P)^2}{8 a^7}
	+\frac{P_{kk} P_{ww}}{a^7}
	+\frac{(Tr P^2)}{4 a^7}
	\bigg)
	\nonumber
	\\
	&
	+\delta q_5^2 
	\bigg(
	\frac{k^2}{4 a^3}
	-\frac{(Tr P)^2}{8 a^7}
	+\frac{P_{vv}P_{ww}}{a^7}
	+\frac{(Tr P^2)}{4 a^7}
	\bigg)
	\nonumber
	\\
	&
	+\delta q_6^2
	\bigg(
	\frac{k^2}{4 a^3}
	+\frac{(P_{vv}+P_{ww})^2}{4a^7}
	-\frac{P_{vw}^2}{a^7}
	-\frac{(Tr P)^2}{8 a^7}
	+\frac{(Tr P^2)}{4 a^7}
	\bigg)
	\nonumber
	\\
	&
	+
\delta q_1\bigg[
	-\frac{\sqrt{2}	P_{kv} \delta \pi_3}{a^3}
	-\frac{\sqrt{2} P_{kw} \delta \pi_4}{a^3}
	-\frac{\sqrt{2} P_{vw} \delta \pi_5}{a^3}
	+\left(\frac{P_{ww}}{\sqrt{2} a^3}
	-\frac{P_{vv}}{\sqrt{2}	a^3}\right) \delta \pi_6
	-\frac{\delta \pi_1 (Tr P)}{6 a^3}
		\nonumber
	\\
	&
	+\delta \pi_2
	\bigg(
	\frac{(Tr P)}{6 a^3}
	-\frac{P_{kk}}{2 a^3}
	\bigg)
	+\delta q_3
	\bigg(
	\frac{\sqrt{2} P_{ww}P_{kv}}{a^7}
	-\frac{7 (Tr P) P_{kv}}{\sqrt{2}	a^7}
	-\frac{\sqrt{2} P_{kw} P_{vw}}{a^7}
	\bigg)
		\nonumber
	\\
	&
	+\delta q_4
	\bigg(
	\frac{\sqrt{2} P_{vv}P_{kw}}{a^7}
	-\frac{7 (Tr P) P_{kw}}{\sqrt{2}	a^7}
	-\frac{\sqrt{2} P_{kv} P_{vw}}{a^7}
	\bigg)
		\nonumber
	\\
	&
	+\delta q_5
	\bigg(
	\frac{\sqrt{2} P_{kk}	P_{vw}}{a^7}
		-\frac{7 	(Tr P)P_{vw}}{\sqrt{2} a^7}
	-\frac{\sqrt{2} P_{kv} P_{kw}}{a^7}
	\bigg)
		\nonumber
	\\
	&
	+\delta q_6
	\bigg(
	-\frac{P_{kv}^2}{\sqrt{2} a^7}
	+\frac{P_{kw}^2}{\sqrt{2}a^7}
	-\frac{(P_{vv}^2-P_{ww}^2)}{\sqrt{2} a^7}
	-\frac{5(P_{vv}-P_{ww}) (Tr P)}{2 \sqrt{2} a^7}
		\bigg)
			\nonumber
		\\
		&
	+\delta q_2 
	\bigg(
	\frac{k^2}{3a^3}
	-\frac{P_{kk}^2}{a^7}
	-\frac{P_{kv}^2}{a^7}
	-\frac{P_{kw}^2}{a^7}
	+\frac{5 (Tr P)^2}{6 a^7}
	-\frac{5 P_{kk} (Tr P)}{2a^7}
	+\frac{(Tr P^2)}{3 a^7}
	\bigg)
	\bigg]
		\nonumber
	\\
	&
	+\delta q_3 \bigg[
	\frac{2 P_{kk} P_{vw}\delta q_4}{a^7}
	+\frac{2 P_{kw} P_{vv}\delta q_5}{a^7}
	+\bigg(\frac{P_{kv} [(TrP)-P_{kk}]}{a^7}
	-\frac{2 P_{kw}	P_{vw}}{a^7}
	\bigg) \delta q_6
		\nonumber
	\\
	&
	-\frac{\sqrt{2}	P_{kv} \delta \pi_1}{3 a^3}
	-\frac{\sqrt{2} P_{kv} \delta\pi_2}{a^3}
	+\frac{P_{vw} \delta \pi_4}{a^3}
	+\frac{2 P_{kw} \delta \pi_5}{a^3}
	+\frac{2 P_{kv} \delta \pi_6}{a^3}
	+\delta \pi_3 \left(\frac{2(Tr P)}{3 a^3}
	-\frac{P_{ww}}{a^3}
	\right)
	\bigg]
		\nonumber
	\\
	&
	+\delta q_4 \bigg[
	\frac{2 P_{kv}P_{ww} \delta q_5}{a^7}
	+\bigg(
	\frac{2P_{kv} P_{vw}}{a^7}
	-\frac{P_{kw} [(TrP)-P_{kk}]}{a^7}
	\bigg)
	\delta q_6
	-\frac{\sqrt{2} P_{kw} \delta \pi_1}{3 a^3}
	-\frac{\sqrt{2}	P_{kw} \delta \pi_2}{a^3}
	+\frac{P_{vw} \delta \pi_3}{a^3}
	\nonumber
	\\
	&
	+\frac{2P_{kv} \delta \pi_5}{a^3}
	-\frac{2 P_{kw} \delta \pi_6}{a^3}
	+\delta\pi_4 \bigg(
	\frac{2 (Tr P)}{3 a^3}
	-\frac{P_{vv}}{a^3}
	\bigg)\bigg]
		\nonumber
	\\
	&
	+\delta q_5
	\bigg[
			\frac{(P_{vv}-P_{ww}) P_{vw}}{a^7}
			 \delta q_6
		-\frac{\sqrt{2} P_{vw} \delta \pi_1}{3a^3}
  		-\frac{\sqrt{2} P_{vw} \delta \pi_2}{a^3}
  		+\delta \pi_5
	\bigg(
	\frac{2 (Tr P)}{3 a^3}
	-\frac{P_{kk}}{a^3}\bigg)\bigg]
	\nonumber
	\\
	&
	+\delta q_6
	\bigg[
		-\frac{P_{vv}-P_{ww}}{3 \sqrt{2} a^3}
 \delta \pi_1
		-\frac{P_{vv}-P_{ww}}{\sqrt{2}a^3}
		 \delta \pi_2
		+\delta \pi_6	\bigg(\frac{2 (Tr P)}{3 a^3}
		-\frac{P_{kk}}{a^3}
		\bigg)\bigg]
			\nonumber
		\\
		&
	+\delta q_2
	\bigg[
	\frac{P_{kk} \delta \pi_2}{a^3}
	+\frac{4 \sqrt{2} P_{kv} \delta \pi_3}{3 a^3}
	+\frac{4 \sqrt{2} P_{kw} \delta \pi_4}{3 a^3}
	-\frac{2 \sqrt{2}P_{vw} \delta \pi_5}{3 a^3}
	-\frac{\sqrt{2} (P_{vv}-P_{ww})}{3	a^3}
     \delta \pi_6
     \nonumber
	\\
	&
	+\delta \pi_1	\left(\frac{(Tr P)}{9 a^3}
	-\frac{P_{kk}}{3 a^3}\right)
	+\delta q_3\bigg(\frac{\sqrt{2} P_{kk} P_{kv}}{3 a^7}
	-\frac{2 \sqrt{2} P_{vv}P_{kv}}{3 a^7}
	+\frac{\sqrt{2} (Tr P) P_{kv}}{3 a^7}
	-\frac{2	\sqrt{2} P_{kw} P_{vw}}{3 a^7}\bigg)
	\nonumber
	\\
	&
	+\delta q_4 \bigg(\frac{\sqrt{2}
		P_{kk} P_{kw}}{3 a^7}-\frac{2 \sqrt{2} P_{ww} P_{kw}}{3
		a^7}+\frac{\sqrt{2} (Tr P) P_{kw}}{3 a^7}-\frac{2 \sqrt{2}
		P_{kv} P_{vw}}{3 a^7}\bigg)
	\nonumber
	\\
	&
	+\delta q_5 
	\bigg(
	\frac{4 \sqrt{2} P_{kv}	P_{kw}}{3 a^7}
	+\frac{\sqrt{2} P_{kk} P_{vw}}{a^7}
	-\frac{5\sqrt{2} P_{vw} (Tr P)}{3 a^7}
	\bigg)
	\nonumber
	\\
	&
	+\delta q_6
	\bigg(
	\frac{2 \sqrt{2} P_{kv}^2}{3 a^7}
	-\frac{2 \sqrt{2} P_{kw}^2}{3a^7}
	-\frac{\sqrt{2} (P_{vv}^2-P_{ww}^2)}{3 a^7}
	-\frac{P_{kk} (P_{vv}-P_{ww})}{\sqrt{2} a^7}
	+\frac{(P_{vv}-P_{ww}) (Tr P)}{3 \sqrt{2}	a^7}
	\bigg)\bigg]
 .
\end{align}
\par
}

\section{Physical Hamiltonian coefficients}\label{coeff}
The background coefficients in the physical Hamiltonian \eqref{HK2}:
\begin{align}
U_1&=\frac{(Tr P)^2+72 \left(P_{kv}^2+P_{kv} P_{kw}+P_{kw}^2+ P_{vw} (P_{kv}+P_{kw})-P_{vw}^2\right)}{36a^4}\nonumber\\
&+\frac{P_{vw}^2 (4 P_{kv}^2 + 4 P_{kw}^2 - 4 P_{vw}^2 - (P_{vv} - P_{ww})^2 -P_{kk}^2)}{a^4P_{kk}^2}-2\frac{\frac{(P_{vv}-P_{ww})^2}{4}}{a^4}\nonumber\\
   &+\frac{54 P_{kk}^4 - 36 P_{kk}^3 (Tr P)}{36a^4P_{kk}^2}+\frac{2 P_{vw} (8 P_{kv} P_{kw} + P_{vw} (Tr P))}{a^4P_{kk}},\nonumber\\
 U_2&=\frac{(Tr P)^2 + 72 (P_{kv}^2 + P_{kv} P_{kw} +P_{kw}^2+ P_{vw}( P_{kv}+P_{kw} )-P_{vw}^2)}{36a^4}\nonumber\\
 &+\frac{\frac{(P_{vv} - P_{ww})^2}{4} (4 P_{kv}^2 + 4 P_{kw}^2 - 4 P_{vw}^2- (P_{vv} - P_{ww})^2-P_{kk}^2)}{a^4P_{kk}^2}-2\frac{P_{vw}^2}{a^4}\nonumber\\
    &+\frac{54 P_{kk}^4 -36 P_{kk}^3(Tr P) }{36a^4P_{kk}^2}+\frac{ 2\frac{(P_{vv} - P_{ww})}{2} (4 P_{kv}^2 - 4 P_{kw}^2 +\frac{(P_{vv} - P_{ww})}{2}(Tr P))}{a^4P_{kk}},\nonumber\\
C_{12}&=\frac{ 
  P_{vw} (P_{vv} - P_{ww}) (4 P_{kv}^2 + 4 P_{kw}^2 - 4 P_{vw}^2 - (P_{vv} - P_{ww})^2)}{a^4P_{kk}^2}+\frac{P_{vw} (P_{vv} - P_{ww}) }{a^4}
 \nonumber
 \\
 &
  \frac{8(P_{kv}^2 - P_{kw}^2)P_{vw}+ 
     8 P_{kv} P_{kw} (P_{vv} - P_{ww}) + 2(Tr P) P_{vw} (P_{vv} - P_{ww})}{a^4P_{kk}}.
\end{align}

\end{document}